\pdfoutput=1
\documentclass[copyright,submission]{eptcs}

\usepackage{nag}
\usepackage[utf8]{inputenc}
\usepackage[T1]{fontenc}
\usepackage{microtype}
\usepackage{amsmath}
\usepackage{amssymb}
\usepackage{amsthm}
\usepackage{mathtools}
\usepackage{color}
\usepackage{tikz}
\usetikzlibrary{positioning, calc, fadings, automata, shapes.multipart}
\usepackage[colorinlistoftodos, textsize=footnotesize]{todonotes}
\usepackage[framemethod=tikz]{mdframed}
\usepackage{tabu}
\usepackage{booktabs}
\usepackage{subcaption}
\usepackage{comment}
\usepackage{adjustbox}
\usepackage{enumitem}
\usepackage{todonotes}
\usepackage{hyperref}
\usepackage[nameinlink, noabbrev, capitalise]{cleveref}

\providecommand{\event}{MARS 2015}

\mdfdefinestyle{thesis}{skipabove=5mm, skipbelow=5mm, frametitlebackgroundcolor=black!20, frametitlerule=true, roundcorner=1.5mm}
\newmdenv[frametitle={Remark}, style=thesis]{remark}

\newcommand{\mi}[1]{\mathit{#1}}

\newcommand{\round}[1]{\left(#1\right)}

\newcommand{\curly}[1]{\left\lbrace#1\right\rbrace}
\newcommand{\abs}[1]{\left|#1\right|}
\newcommand{\set}[1]{\curly{#1}}

\crefname{version}{version}{versions}

\tikzset{force/.style={thick, ->}}
\tikzset{initial text={}}

\newenvironment{hioa}{
    \begin{adjustbox}{scale=0.75}\begin{tikzpicture}[auto, shorten >=1pt, shorten <=1pt, node distance=3cm, on grid, >=stealth, every state/.style={rectangle, rounded corners=2mm, rectangle split, rectangle split parts=3}]
}{
    \end{tikzpicture}\end{adjustbox}
}

\newcommand{\shortalign}{
    \setlength{\abovedisplayskip}{0pt}
    \setlength{\belowdisplayskip}{0pt}
    \setlength{\abovedisplayshortskip}{0pt}
    \setlength{\belowdisplayshortskip}{0pt}
}

\newcommand{\surface}{
    \begin{scope}[transparency group]
        \clip (-2,0) rectangle (2,-1);
        \draw[scope fading=south] (-2,0) rectangle (2,-1);
        \foreach \x in {-2,-1.75,...,3}
            \draw (\x,0) -- ($(\x,0) - (1,1)$);
    \end{scope}
}
\newcommand{\wheel}{
    \draw (0, 1) circle (1);
    \draw[fill] (0, 1) circle (0.05);
}
\newcommand{\saddle}[1][110]{
    \draw (0, 1) -- ([shift=(#1:3)] 0,1);
    \draw[fill] ([shift=(#1:0)] 0,1) circle (0.05);
    \draw[fill] ([shift=(#1:2)] 0,1) circle (0.03);
    \draw[fill] ([shift=(#1:3)] 0,1) circle (0.05);
}
\newcommand{\angletheta}[1][110]{
    \draw[dashed] (0, 1) -- (0, 3);
    \draw[dashed] ([shift=(90:1.5)] 0,1) arc (90:#1:1.5) node [midway, below] {$\theta$};
}

\AtBeginDocument
{
    \setlength\abovedisplayskip{.3\baselineskip plus .1\baselineskip}
    \setlength\abovedisplayshortskip{.3\baselineskip plus .1\baselineskip}
    \setlength\belowdisplayskip{.2\baselineskip plus .1\baselineskip}
    \setlength\belowdisplayshortskip{.2\baselineskip plus .1\baselineskip}
}

\setlist[itemize]{noitemsep}
\setlist[enumerate]{noitemsep}

\title{On the Control of Self-Balancing Unicycles}

\author{Felix Freiberger \qquad Holger Hermanns
    \institute{Saarland University -- Computer Science}
}

\def\titlerunning{On the Control of Self-Balancing Unicycles}
\def\authorrunning{Felix Freiberger, Holger Hermanns}

\begin{document}

\maketitle

\begin{abstract}
    This paper discusses the problem of designing a self-balancing unicycle where pedals are used for both power generation and speed control. After developing the principal physical aspects (in the longitudinal dimension), we describe an abstract model in the form of a collection of hybrid automata, together with design requirements to be met by an ideal controller. We discuss simplifications and assumptions that make this model amenable to verification and validation tools such as SpaceEx. To enable experimentation with different prototypical controllers and user behaviours in concrete scenarios, we also develop a simple simulation framework using digital time.
\end{abstract}

\section{Introduction}
 During the last decade, self-balancing devices for individual
 (short-distance) travel have gained popularity. The most prominent
 example is the \href{http://www.segway.com/}{Segway}, a two wheeler
 with a handrail for the user to hold on to, but there are also
 devices that require the user to stand freely on a single wheel, such
 as the \href{http://www.solowheel.eu/}{Solowheel}. In both designs
 the tilting of the foot rest is continuously taken as sensor input,
 and interpreted as the intention to accelarate or slow
 down. Accelaration is thus the result of leaning forwards (or pushing
 the handrail, or tilting the feet downwards), and conversely for
 deacceleration.

For classical pedal-powered unicylces the task of speed control and of
balancing is entirely left to the rider. In this work, we analyze
self-balancing unicycles which structurally resemble a normal
unicycle, but are characterised by the following three properties:

\begin{enumerate}
\item
    The unicycle needs to be balanced both longitudinally and laterally.
\item
    As with a normal cycle, the rider influences the speed by pedaling,
    not by leaning forwards or backwards.
\item
The seat post is to be kept stably upright, thus perpendicular to the ground horizon. 
\end{enumerate}

In particular, this implies that the rider does effectively not lean
forwards or backwards independent of the device's wheel rotation since
he has a fixed seat position.

We are looking for a controller that enables the above features. Our
design is meant to echo the concept of a pedelec, a pedal-assisted
electric two-wheel bicycle. A pedelec supports the rider by an
electric motor, which emits additional power to one of the rotating
wheels proportional to the measured pedalling activity. The electric
power is drawn from an on-board battery. Pedelecs are rapidly gaining
popularity in many countries, especially across central Europe.

We assume a unicylce construction where the pedals are mounted to a
generator, providing electric power to the system and at the same time
indicating the current intended riding speed to the controller. The
pedals are thus \emph{not} connected to the wheel. This is because
otherwise the needed corrective torques will have to interfere with
the pedalling feet, making it presumably very inconvenient to ride
smoothly.  Instead, a hub motor is mounted between the wheel and the
rods the saddle is mounted on. This motor can therefore enact a torque
between the wheel and the saddle, and thus the rider.

This paper develops the principal physical aspects focussing on the
longitudinal dimension, thus assuming that it is the rider's
responsibility not to fall over laterally (just as when riding an
ordinary bicycle). We represent the system as a collection of hybrid
automata and discuss design requirements to be met by an ideal
controller. We then turn our attention to simplifying assumptions so
as to make this model amenable to verification and validation tools
for affine dynamics. In this we mainly target
SpaceEx~\cite{DBLP:conf/cav/FrehseGDCRLRGDM11}, one of the most
advanced verification environments in this context.
We also discuss briefly how
prototypical controller instances can be experimented with on the
abstract model, making use of a simple simulation framework using
digital time.

To the best of our knowledge, the problem of designing a
pedal-assisted electric unicycle is in itself novel, yet it is very
intriguing. Solving it will mean the conception of the lightest
pedelec ever built, and this might come with commercialisation chances
for commuter traffic in urban areas, for instance. Related work covers
the classical inverted pendulum~\cite{DBLP:journals/robotica/FengY88},
as well as Segway-style designs~\cite{chessway}.

The paper is organized as follows.
In \cref{sect:model-aspects}, we discuss the physical aspects of such a unicycle. \cref{sect:hioa} develops the hybrid automata model, and  \cref{sect:controller-design} discusses  controller designs for it.
In \cref{sect:verification}, we lay out our verification efforts, and \cref{sect:simulation} briefly reviews a simulation framework for our model. Finally, in \cref{sect:conclusion}, we conclude.

\section{Aspects of the Model}\label{sect:model-aspects}

\subsection{A Unicycle}\label{sect:physics-summary}

\begin{figure}
    \centering\includegraphics[scale=0.5]{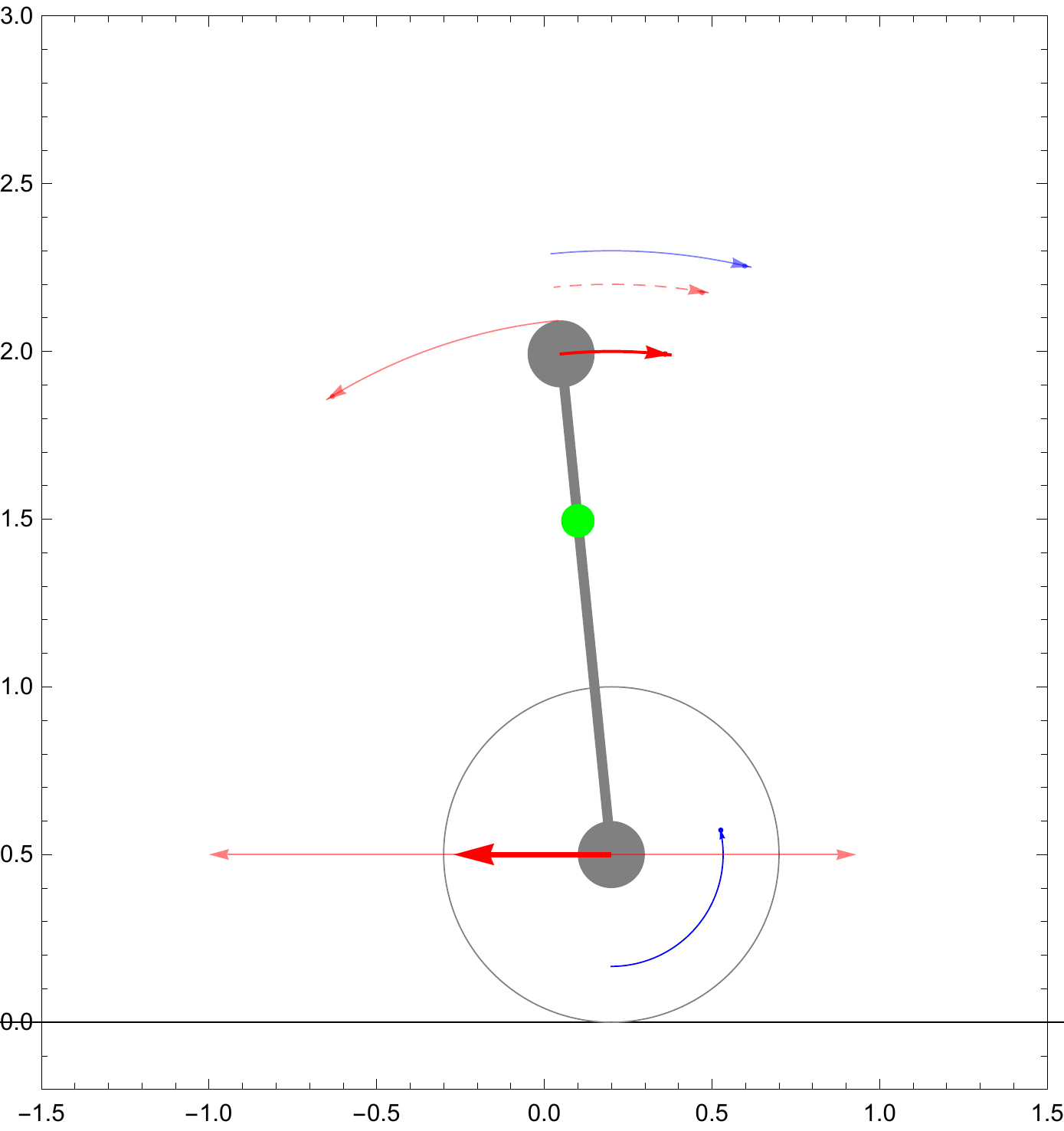}
    \caption[A simplified unicycle with forces]{\label{fig:system-forces} \centering A Simplified Unicycle With Forces \par The blue arrows both represent $\tau_{W, \mi{mot}}$. The thin red arrows on the wheel represent $F_{W, \mi{mot}}$ (to the left) and $F_{W, g}$ (to the right), while the thick one is $F_W$. The red dashed torque on the rod is $\tau_{S, W}$, the thin, solid one shows $\tau_{S, g}$. Here, too, the thick arrow represents the sum, $\tau_{S}$.}
\end{figure}

\paragraph{Basic Setup.}
We begin by discussing the behaviour of a pure unicycle.
Our model of a unicycle is sketched in \cref{fig:system-forces}. At the
bottom, there is a large wheel $W$. In its centre, a long rod is
affixed which has the saddle mounted on its top. The rider sits on the
saddle. We assume that rod, saddle and rider do not move relative to
each other and call the resulting object \emph{saddle} ($S$).

We consider only two external forces influencing this system:
\begin{enumerate}
\item gravity, pulling down on $S$, and
\item a torque $\tau_{W, \mi{mot}}$ applied between $W$ and $S$, caused by a motor mounted there (where a positive torque rotates $W$ counter-clockwise and $S$ clockwise).
\end{enumerate}

To simplify the model, we assume the following:
\begin{enumerate}
\item There are no lateral forces, so the forces we consider live in a two-dimensional vector space.
\item The friction between $W$ and the ground is high enough for the wheel to never slide.
\end{enumerate}

Then, the model can be described by the parameters specified
in \cref{tbl:model-parameters}. A state of the model is represented by
the variables in \cref{tbl:model-state}.

\begin{table}
     \centering\begin{tabu} to 0.9\linewidth {X[1]X[5]}
        \toprule
        \rowfont{\bfseries} parameter & description \\\midrule
        $r_W$ & the radius of wheel $W$ \\
        $r_\mi{com}$ & the distance from the bottom of $S$ to its center of mass \\
        $r_S$ & the distance from the bottom of $S$ to the top \\
        $m_{W, \mi{real}}$ & the actual mass of $W$ \\
        $m_S$ & the mass of $S$, including the driver's mass \\
        $I_S$ & the moment of inertia of $S$ \\
        $I_W$ & the moment of inertia of $W$ \\
        $\xi$ & constant with $\xi \in [0.5, 1]$ describing the distribution of the mass in the wheel, see \cref{sect:abstractinginertia} \\
        $\tau_\mi{max}$ & the maximum torque the motor can enact in any direction \\
        \bottomrule
    \end{tabu}
    \caption{\label{tbl:model-parameters} Parameters of the System}
\end{table}

\begin{table}
     \centering\begin{tabu} to 0.9\linewidth {X[1.5]X[5]}
        \toprule
        \rowfont{\bfseries} state variable & description \\\midrule
        $x_W$ & the $x$-coordinate of the wheel \\
        $v_W$ & the horizontal speed of the wheel \\
        $\theta$ & the angle of the rod, relative to its upright position \\
        $\omega_S$ & the angular speed of the rod \\
        \bottomrule
    \end{tabu}
    \caption{\label{tbl:model-state} State of the System}
\end{table}

\paragraph{Equations of Motion.}

We now briefly summarize the equations of motion. Their full
derivation can be found in \cref{sect:physics}.

The torque $\tau_{W, \mi{mot}}$ rotates the wheel $W$ counter-clockwise, which corresponds to a
force $F_{W, \mi{mot}}$ that pulls $W$ to the right with
\[ F_{W, \mi{mot}} = -\frac{-\tau_{W, \mi{mot}}}{r_W} . \tag{\ref{eqn:defining-f-w-d}} \]
Gravity pulls down on $S$. If it is not in the upright position, this causes a counter-clockwise $\tau_{S, g}$ with
\[ \tau_{S, g} = m_s \cdot g \cdot \sin \theta \cdot r_\mi{com} \tag{\ref{eqn:defining-tau-s-g}} \]
and a force $F_{W, g}$ that pushes $W$ to the right with
\[ F_{W, g} = m_s \cdot g \cdot \cos \theta \cdot \sin \theta . \tag{\ref{eqn:defining-f-w-g}} \]
We now have collected all longitudinal forces that act on $W$, we call their sum $F_W$:
\[ F_W = F_{W, \mi{mot}} + F_{W, g} \tag{\ref{eqn:defining-f-w}} \]
This force acts on $W$, which is affixed to $S$. To analyse the impact of this force on the system, we split it into two parts $F_{W,1} + F_{W,2} = F_W$, where $F_{W,1}$ and $F_{W,2}$ can be seen as acting on $S$ in isolation and the wheel $W$ in isolation, respectively:
\begin{align}
    m_W &= m_{W, \mi{real}} + \xi \cdot m_{W, \mi{real}} \tag{\ref{eqn:adjusted-mass}} \\
    \beta &= \frac{m_S}{m_W} \cdot \round{1 + 3 \cos^2 \theta} \tag{\ref{eqn:defining-beta}} \\
    F_{W,1} &= F_W \cdot \frac{\beta}{1 + \beta} \tag{\ref{defining-f-w-1}} \\
    F_{W,2} &= F_W \cdot \frac{1}{1 + \beta} \tag{\ref{defining-f-w-2}}
\end{align}
While $F_{W,2}$ can now be seen as the only force affecting the longitudinal movement of the unicycle (or more specifically its wheel), $F_{W,1}$ has a more complicated influence: It causes a counter-clockwise torque $\tau_{S,W}$ on $S$ with strength
\[ \tau_{S,W} = F_{W,1} \cdot r_\mi{com} \cdot \cos \theta . \tag{\ref{defining-tau-s-w}} \]
We now know all torques acting on $S$:
\[ \tau_S = -\tau_{W, \mi{mot}} + \tau_{S, g} + \tau_{S, W} . \tag{\ref{eqn:defining-tau-s}} \]

Having considered all forces and torques, we can now give differential equations for the $x$-position, speed, angle and angular velocity of the unicycle:
\begin{align}
\dot{x_W} &= v_W \tag{\ref{eqn:flow-xw}} \\
\dot{v_W} &= \frac{F_{W, 2}}{m_W} \tag{\ref{eqn:flow-vw}} \\
\dot{\theta} &= \omega \tag{\ref{eqn:flow-theta}} \\
\dot{\omega} &= \frac{\tau_S}{I_S} \tag{\ref{eqn:flow-omega}}
\end{align}

\Cref{fig:system-forces} displays an exemplary state of the unicycle where all linear forces and torques are indicated.

\subsection{Surrounding Components}

In addition to the unicycle, we need to add three additional
components to the model:  user,  controller and motor.

Since we are considering a device where the pedals generate
electric power and indicate the intended speed, the
user has no other influence on the actual unicycle. Therefore, we
model the user as a black-box component that sets a variable
$\mi{userIntent}$, indicating the target speed.

The controller may use the unicycle's state variables and
$\mi{userIntent}$ to compute the torque the motor should enact, and
sets this as $\mi{motorTorque}$. We only consider controllers
where the development of $\mi{motorTorque}$ over time is
continuous and differentiable.

Finally, the motor is a  model of  the physical constraints of the
motor's capabilities. It uses $\mi{motorTorque}$ to set the actual
force $\tau_{W, \mi{mot}}$ applied between saddle and wheel. In our model,
this is simply $\mi{motorTorque}$ capped to the maximal strength of
the motor, which we denote by $\tau_\mi{max}$.

\section{A Hybrid Automaton Model}
\label{sect:hioa}

In the following, we consider a fixed unicycle and therefore consider
the parameters specified in \cref{tbl:model-parameters} and
$\tau_\mi{max}$ to be constants.  We use the hybrid (input/output) automata (HIOA)
formalism~\cite{DonzeF13} due to its conciseness and adequacy. We
assume familiarity with this notation, which we shall use in its
intuitive pictorial form below.

\begin{figure}[t]
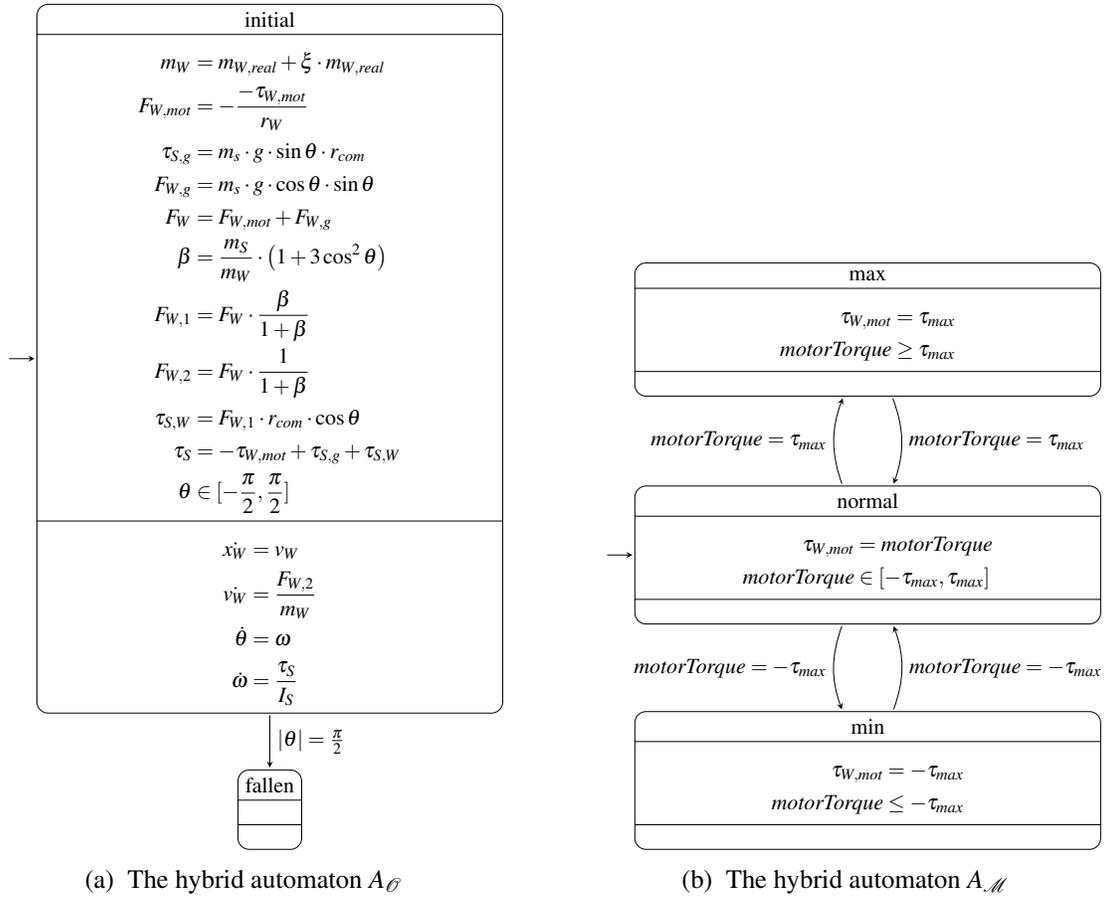

    \subcaptionbox{\label{fig:unicycle-hioa} The hybrid automaton $A_\mathcal{O}$}[0.5\linewidth]{
        \centering\begin{hioa}\tikzset{node distance=8cm}
            \node[state, initial] (always) {
                initial
                \nodepart{two}
                \begin{minipage}{0.5\textwidth}\shortalign
                    \begin{align*}
                        m_W &= m_{W, \mi{real}} + \xi \cdot m_{W, \mi{real}} \\
                        F_{W, \mi{mot}} &= -\frac{-\tau_{W, \mi{mot}}}{r_W} \\
                        \tau_{S, g} &= m_s \cdot g \cdot \sin \theta \cdot r_\mi{com} \\
                        F_{W, g} &= m_s \cdot g \cdot \cos \theta \cdot \sin \theta \\
                        F_W &= F_{W, \mi{mot}} + F_{W, g} \\
                        \beta &= \frac{m_S}{m_W} \cdot \round{1 + 3 \cos^2 \theta} \\
                        F_{W,1} &= F_W \cdot \frac{\beta}{1 + \beta} \\
                        F_{W,2} &= F_W \cdot \frac{1}{1 + \beta} \\
                        \tau_{S,W} &= F_{W,1} \cdot r_\mi{com} \cdot \cos \theta \\
                        \tau_S &= -\tau_{W, \mi{mot}} + \tau_{S, g} + \tau_{S, W} \\
                        \theta &\in [-\frac{\pi}{2}, \frac{\pi}{2}]
                    \end{align*}
                \end{minipage}
                \nodepart{three}
                \begin{minipage}{0.5\textwidth}\shortalign
                    \begin{align*}
                        \dot{x_W} &= v_W \\
                        \dot{v_W} &= \frac{F_{W, 2}}{m_W} \\
                        \dot{\theta} &= \omega \\
                        \dot{\omega} &= \frac{\tau_S}{I_S}
                    \end{align*}
                \end{minipage}
            };
            \node[state, below of=always] (fallen) {fallen};
            \draw (always) edge [->] node {$\abs{\theta} = \frac{\pi}{2}$} (fallen);
        \end{hioa}
    }%
    \subcaptionbox{\label{fig:transmission-hioa} The hybrid automaton $A_\mathcal{M}$}[0.5\linewidth]{
        \centering\begin{hioa}\tikzset{node distance=4cm}
            \node[state, initial] (normal) {
                normal
                \nodepart{two}
                \begin{minipage}{0.5\textwidth}\shortalign
                    \begin{align*}
                        \tau_{W, \mi{mot}} &= \mi{motorTorque} \\
                        \mi{motorTorque} &\in [-\tau_\mi{max}, \tau_\mi{max}]
                    \end{align*}
                \end{minipage}
            };
            \node[state, above of = normal] (max) {
                max
                \nodepart{two}
                \begin{minipage}{0.5\textwidth}\shortalign
                    \begin{align*}
                        \tau_{W, \mi{mot}} &= \tau_\mi{max} \\
                        \mi{motorTorque} &\ge \tau_\mi{max}
                    \end{align*}
                \end{minipage}
            };
            \node[state, below of = normal] (min) {
                min
                \nodepart{two}
                \begin{minipage}{0.5\textwidth}\shortalign
                    \begin{align*}
                        \tau_{W, \mi{mot}} &= -\tau_\mi{max} \\
                        \mi{motorTorque} &\le -\tau_\mi{max}
                    \end{align*}
                \end{minipage}
            };
            \draw (normal) edge [->, bend left = 20] node {$\mi{motorTorque} = \tau_\mi{max}$} (max);
            \draw (normal) edge [->, bend right = 20] node [swap] {$\mi{motorTorque} = -\tau_\mi{max}$} (min);
            \draw (max) edge [->, bend left = 20] node {$\mi{motorTorque} = \tau_\mi{max}$} (normal);
            \draw (min) edge [->, bend right = 20] node [swap] {$\mi{motorTorque} = -\tau_\mi{max}$} (normal);
        \end{hioa}
    }
    \caption{\label{fig:hioa} Graph representations of our model components}
\end{figure}

\begin{sloppypar}
\paragraph{Unicycle.}
The unicycle behaviour is represented as the HIOA
$\mathcal{O} = (A_\mathcal{O}, C_\mathcal{O}, O_\mathcal{O})$ over variables $X_\mathcal{O} = \set{\tau_{W, \mi{mot}}, x_W, v_W, \theta, \omega}$ with hybrid automaton \cite{DBLP:journals/tcs/AlurCHHHNOSY95} $A_\mathcal{O}$ as depicted in \cref{fig:unicycle-hioa}, controlled variables $C_\mathcal{O} = \set{x_W, v_W, \theta, \omega}$, output variables $O_\mathcal{O} = C_\mathcal{O}$ and where initially $x_W = v_W = \theta = \omega = 0$. For clarity, we use temporary variables that are not in $X_\mathcal{O}$ and can be removed by substitution within invariants. The location `fallen' represents that the saddle is on the ground.
\end{sloppypar}

\paragraph{Motor.}
We also formalize the motor constraints: Let $\mathcal{M} = (A_\mathcal{M}, C_\mathcal{M}, O_\mathcal{M})$ be a hybrid I/O automaton over the variable $X_\mathcal{M} = \set{\mi{motorTorque}, \tau_{W, \mi{mot}}}$ with $A_\mathcal{M}$ as defined by \cref{fig:transmission-hioa} with $C_\mathcal{M} = \emptyset$ and $O_\mathcal{M} = \set{\tau_{W, \mi{mot}}}$.

\paragraph{Full model.}
For the user and controller, we assume abstract models given by hybrid I/O automata $\mathcal{U}$ and $\mathcal{C}$ which are thus far left unspecified. Then, the full model can be obtained by parallel composition of the HIOA~\cite{DonzeF13,DBLP:journals/tcs/AlurCHHHNOSY95}:
\[\mathcal{O} \parallel \mathcal{M} \parallel \mathcal{U} \parallel \mathcal{C}\]

In the above, we have not detailed the user and the controller
behaviour. For the user $\mathcal{U}$, who in our model simply
controls the pedalling speed, this is intentional, since conceptually
the design should tolerate any user behaviour. So, the pedalling speed
can be viewed as a continuous input $\mi{driveIntent}$ to the system.

\section{Controller design}\label{sect:controller-design}

We now to turn to the design of a controller $\mathcal{C}$ for the
unicylce.

\paragraph{Constraints on the Controller Design.}
 For any given user, an optimal controller for this system
has the following goals:

\begin{enumerate}
\item
    The unicycle should normally be upright, i.\,e. whenever $\mi{driveIntent}$ remains constant, $\theta$ should asymptotically approach $0$.
    Crucially, independent of user input, the unicycle may never enter state `fallen', i.\,e. $\theta$ may never leave the range $]-\frac{\pi}{2}, \frac{\pi}{2}[$.

\item
    The unicycle should adhere to the speed set by the user, i.\,e. whenever $\mi{driveIntent}$ remains constant, $v_W$ should approach $\mi{driveIntent}$.
\end{enumerate}

\paragraph{A Simple Controller.}
The design space for a valid controller is enormous. Unfortunately, we
are not aware of tool support for constructing the optimal controller,
even though this is an active research area in general. We instead
look into a family of relatively simple yet widely applicable controllers,
namely proportional-integral-derivative (PID) controllers.
PID controllers only try to minimize a single
error term. In our model, however, there are two (possibly
conflicting) goals to optimize for. A pragmatic way of solving this
uses a weighted sum as the error term:
\[ \mi{error} \coloneqq \theta + (\mi{driveIntent} - v_W) \cdot \mi{intentFactor}
\]
Here, $\mi{intentFactor}$ is a parameter of the model indicating how
much to prioritize achieving the desired speed over staying upright.

\section{Towards Model Verification}\label{sect:verification}

Being a HIOA, our formal model appears to lend itself to analysis with
SpaceEx~\cite{DBLP:conf/cav/FrehseGDCRLRGDM11}. However, there are
multiple issues that need to be addressed.

\paragraph{Bounding.}
The model contains many variables which are unbounded, preventing direct analysis by SpaceEx. This affects for example the horizontal position $x_W$ and the angular velocity $\omega$.

While some variables are not crucial for the analysis and can be removed, such as $x_W$, others are an integral part of the model. The angular velocity $\omega$ influences the angle $\theta$, which is relevant for both further movement and part of the goals of the controller; making it impossible to eliminate $\omega$ from the model.
However, it is possible to introduce artificial bounds that do not affect analysis: It is safe to consider the unicycle as fallen when $\omega$ leaves an interval of plausible values because the physical constraints prevent the controller from stopping the fall soon enough.

\paragraph{Linearisation.}

\begin{figure}
    \subcaptionbox{horizontal force on the wheel}[0.5\linewidth]{
        \centering\includegraphics[scale=0.5]{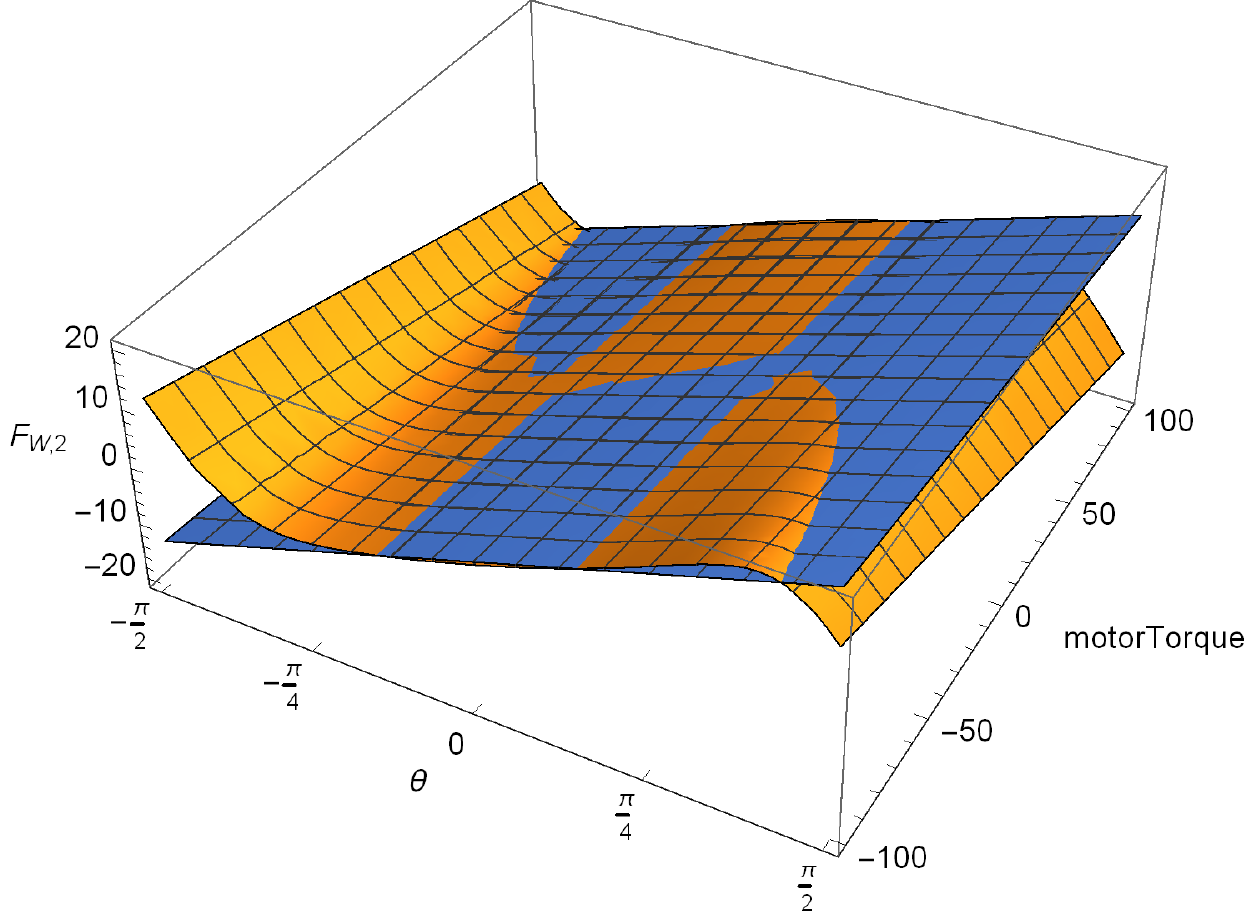}
    }%
    \subcaptionbox{torque on the saddle}[0.5\linewidth]{
        \centering\includegraphics[scale=0.5]{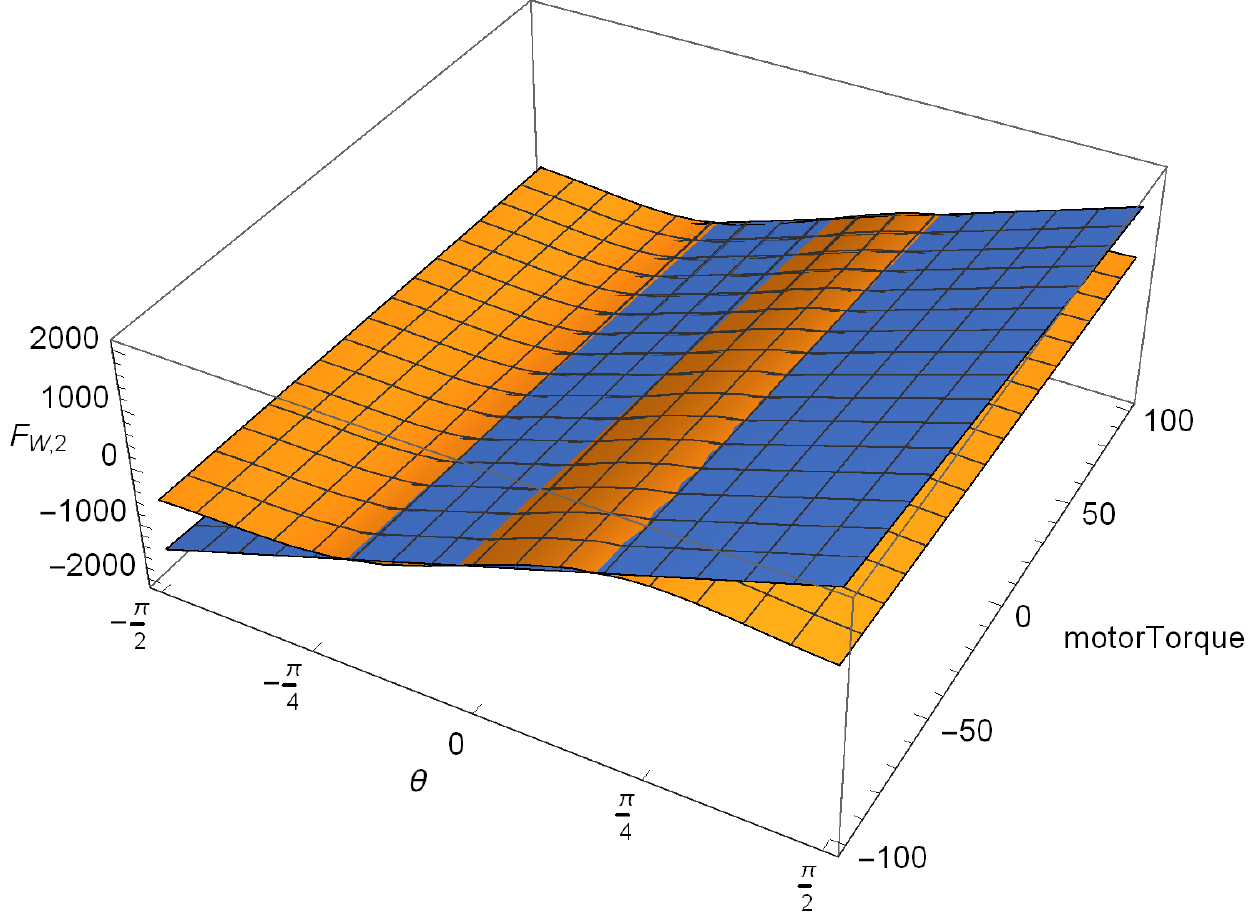}
    }
    \caption{\label{fig:linearization} The dependency of the physical forces on the angle $\theta$ and the motor's force $\mi{motorTorque}$ for a model with exemplary parameters (orange) and a linearised version thereof (blue).}
\end{figure}

The behaviour of the system is non-linear as apparent from the equations in \cref{fig:unicycle-hioa}. \Cref{fig:linearization} depicts the situation for exemplary, realistic parameters. We numerically obtained linearisations, optimized for $\theta \in [-\frac{\pi}{4}, \frac{\pi}{4}]$. They are depicted in blue.

SpaceEx can only analyse systems with linear behaviour~\cite{DBLP:conf/cav/FrehseGDCRLRGDM11}. We conclude that while our full model cannot be analysed with it, there are linearised versions with only reasonable errors within the range of angles $\theta$ that is most relevant to analysis.

\section{Simulation}\label{sect:simulation}

To allow easy experimentation with different approaches for controller design without having to model them as an HIOA that is amenable to formal analysis, we developed a simulation framework in Mathematica. It approximates the model's behaviour for a concrete controller and user input using digital time.

\begin{figure}
    \centering\begin{tikzpicture}[scale=0.55, every node/.style={transform shape}]
        \tikzset{s/.style={draw, rectangle, minimum width = 7cm, minimum height=1cm, align = center}, e/.style={draw, ->, shorten >=1pt, shorten <=1pt}, node distance = 1.5cm, auto, font=\large}
        \path[fill=black!10] (-6.5, 1) rectangle (4, -11.5);
        \begin{scope}
            \node[s] at (0, 0) (s1) {advance time by $\Delta t$};
            \node[s, below of = s1] (s2) {compute $F_{W, 2}$ and $\tau_S$};
            \node[s, below of = s2] (s3) {update $v_W$, $x_W$, $\omega$ and $\theta$ \\ assuming constant forces for $\Delta t$ time};
            \node[s, below of = s3] (s4) {compute observations \\ visible to the controller};
            \node[s, below of = s4] (s5) {evaluate the user input};
            \node[s, below of = s5] (s6) {evaluate the controller};
            \node[s, below of = s6] (s7) {update $\mi{motorTorque}$};
            \node[s, below of = s7] (s8) {create log entry};
            \draw[e] (s1) to (s2);
            \draw[e] (s2) to (s3);
            \draw[e] (s3) to (s4);
            \draw[e] (s4) to (s5);
            \draw[e] (s5) to (s6);
            \draw[e] (s6) to (s7);
            \draw[e] (s7) to (s8);
            \draw[e] (s8.west) to ($(s8.west) - (1, 0)$) to node [align=right] {fixed \\ iteration \\ count} ($(s1.west) - (1, 0)$) to (s1.west);
        \end{scope}
        \begin{scope}
            \node[s] at (10, -5) (plot) {plot log as graph};
            \node[s] at (10, -7) (animate) {render log as animation};
            \draw[e] (s8.east) to ($(s8.east) + (2, 0)$) |- (plot.west);
            \draw[e] (s8.east) to ($(s8.east) + (2, 0)$) |- (animate.west);
        \end{scope}
    \end{tikzpicture}
    \caption{\label{fig:simulation-algorithm} The order of evaluation in the simulation framework.}
\end{figure}
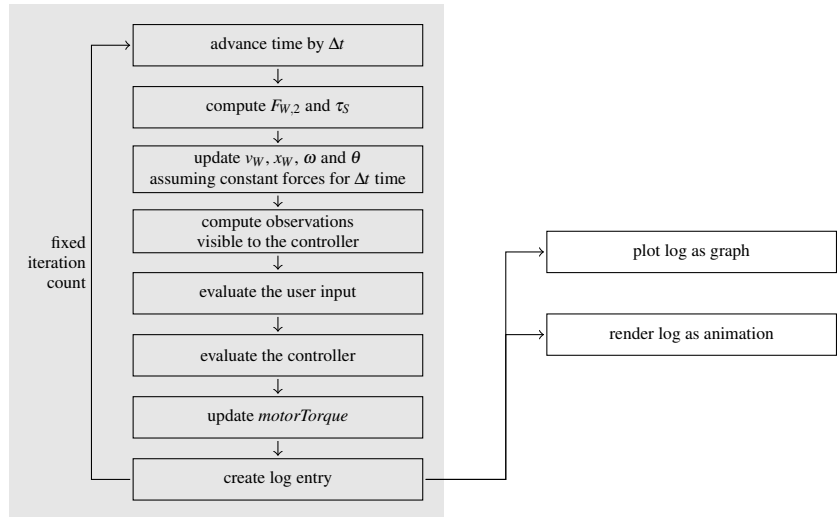

\Cref{fig:simulation-algorithm} shows the core of the simulation algorithm. To visualize the result, the framework can generate graphs plotting certain model parameters as well as animations displaying the state of the unicycle and the forces acting on it over time. \Cref{fig:system-forces} has been generated using this framework.

\section{Conclusion}\label{sect:conclusion}
This paper has developed a model of a unicycle where pedals are used
for both power generation and speed control. We have discussed
simplifications needed to make the model amenable to verification, and
have presented a simple simulation environment.

\paragraph{Additional Resources.} Additional materials such as model files and the simulation framework can be found at \mbox{\url{http://depend.cs.uni-saarland.de/~freiberger/unicycles/}}, also including animations and an implementation of a PID-based controller.

\paragraph{\textit{Acknowledgements.}} The authors are grateful to  Felix Maurer for  double-checking the physics, and to the contributors to the discussion \href{http://physics.stackexchange.com/questions/179622/movement-and-rotation-of-a-inverted-pendulum-like-object-with-external-forces}{``Movement and Rotation of a Inverted-Pendulum-like Object with External Forces''}.

\bibliography{bibliography}
\bibliographystyle{eptcs}

\clearpage\appendix

\section{Physics of a Unicycle}\label{sect:physics}

In this appendix, we justify and explain the derivations in \cref{sect:physics-summary}.

\subsection{Abstracting Away the Moment of Inertia of the Wheel}\label{sect:abstractinginertia}

For a moment, let us just consider the wheel $W$. If it is moving, it is both rotating around its center and translating along the ground. If a horizontal force $F$ is acting on it, its effect is governed by both the inertia and moment of inertia of $W$.

We will now derive the movement equations as in~\cite{halliday2013fundamentals} for $W$ under such an external force $F$, yielding a way to simplify the model.

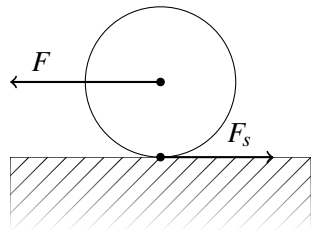
\begin{figure}[b]
    \centering\begin{tikzpicture}
        \surface
        \wheel
        \draw[force] (0, 1) -- node [above left = 0mm and 3mm] {$F$} (-2, 1);
        \draw[fill] (0, 0) circle (0.05);
        \draw[force] (0, 0) -- node [above right] {$F_s$} (1.5, 0);
    \end{tikzpicture}
    \caption{\label{fig:wheel-with-force} Wheel $W$ with an External Force $F$}
\end{figure}

The system is sketched in \cref{fig:wheel-with-force}. The external force $F$ affects the center of mass of the wheel. The static friction between the wheel and the ground causes a force $F_s$ that is pulling the bottom of the wheel against the direction of $F$.

We will call the linear acceleration of $W$ to the left $a$, and the counterclockwise angular acceleration $\alpha$. Therefore, we have:
\begin{align}
    \label{eqn:force-acceleration} F - F_s &= m_{W, \mi{real}} \cdot a \\
    \label{eqn:torque-acceleration} F_s \cdot r_W &= I_W \cdot \alpha
\end{align}

Since the wheel is not slipping, we can relate $a$ and $\alpha$:
\begin{equation}\label{eqn:acceleration-alpha}
    \alpha = \frac{a}{r_W}
\end{equation}

By substituting \cref{eqn:acceleration-alpha} in \cref{eqn:torque-acceleration}, we obtain:
\begin{equation}
    F_s = I_W \cdot \frac{a}{r_W ^2}
\end{equation}

By substituting this in \cref{eqn:force-acceleration}, we obtain:
\begin{equation}\label{eqn:acceleration-inertia}
    a = \frac{F}{m_{W, \mi{real}} + \frac{I_W}{r_W ^2}}
\end{equation}

The moment of inertia $I$ depends on the mass distribution in $W$. If it is distributed evenly in a cylindrical wheel, we have $I = \frac{1}{2} \cdot m_{W, \mi{real}} \cdot r_W ^2$; if the mass is only in the outermost points (effectively making the wheel a ring), we have $I = m_{W, \mi{real}} \cdot r_W ^2$. In reality, the mass distribution will be somewhere in between, leading to
\begin{equation}\label{eqn:moment-of-inertia}
    I = \xi \cdot m_{W, \mi{real}} \cdot r_W ^2
\end{equation}
for a $\xi \in [0.5, 1]$.

By substituting \cref{eqn:moment-of-inertia} in \cref{eqn:acceleration-inertia}, we finally obtain:
\begin{equation}
    a = \frac{F}{m_{W, \mi{real}} + \xi \cdot m_{W, \mi{real}}}
\end{equation}

Coincidentally, this equation also describes the frictionless acceleration of an object with the mass
\begin{equation}\label{eqn:adjusted-mass}
    m_W \coloneqq m_{W, \mi{real}} + \xi \cdot m_{W, \mi{real}} .
\end{equation}

In the following sections, we will therefore neglect any rotational forces on $W$ and use the \emph{adjusted mass} $m_W$ to accommodate them instead.

\subsection{From External Torque to a Linear Force}

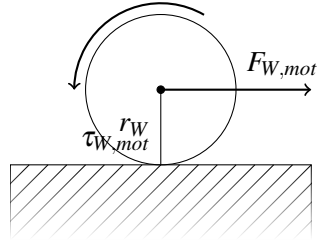
\begin{figure}
    \centering\begin{tikzpicture}
        \surface
        \wheel
        \draw[force] ([shift=(60:1.15)]0, 1) arc(60:180:1.15) node [midway, above left] {$\tau_{W, \mi{mot}}$};
        \draw[force] (0, 1) -- node [above right] {$F_{W, \mi{mot}}$} (2, 1);
        \draw (0, 1) -- node [left] {$r_W$} (0, 0);
    \end{tikzpicture}
    \caption{\label{fig:torque-to-force} Converting $\tau_{W, \mi{mot}}$ to $F_{W, \mi{mot}}$}
\end{figure}

We can now think of the unicycle as a point of mass $W$ that slides frictionless along a surface, with the rod $S$ mounted on it. Since the external torque $\tau_{W, \mi{mot}}$ is rotating $W$ counterclockwise using a moment arm of length $r_W$, we can view it as a force $F_{W, \mi{mot}}$ that is pulling $W$ to the right with
\begin{equation}\label{eqn:defining-f-w-d}
    F_{W, \mi{mot}} = -\frac{-\tau_{W, \mi{mot}}}{r_W},
\end{equation}
as shown in \cref{fig:torque-to-force} (where $F_{W, \mi{mot}}$ is \emph{negative} and therefore points to the left).

\subsection{Adding Gravity}

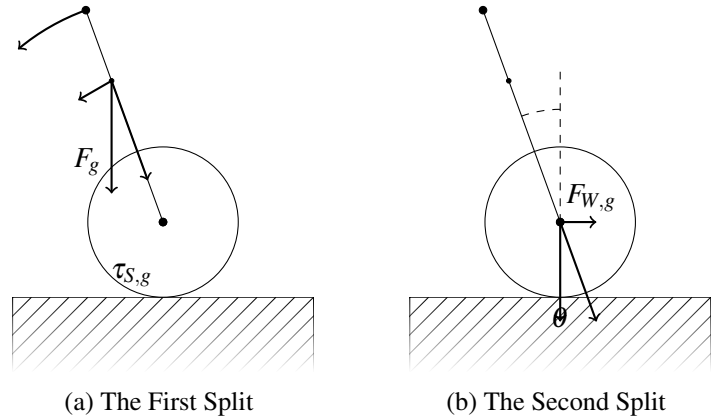
\begin{figure}\centering
    \subcaptionbox{The First Split}[0.33\linewidth]{
        \centering\begin{tikzpicture}
            \surface
            \wheel
            \saddle
            \draw[force] ([shift=(110:2)] 0,1) -- node [below left] {$F_g$} ([shift=(110:2)] 0, -0.5);
            \draw[force] ([shift=(110:2)] 0,1) -- ([shift={($(110:2) - cos(20)*(110:1.5)$)}] 0,1);
            \draw[force] ([shift=(110:2)] 0,1) -- ([shift={($(110:2) + sin(20)*(210:1.5)$)}] 0,1);
            \draw[force] ([shift=(110:3)] 0,1) arc (110:130:3) node [midway, above left] {$\tau_{S, g}$};
        \end{tikzpicture}
    }\subcaptionbox{The Second Split}[0.33\linewidth]{
        \centering\begin{tikzpicture}
            \surface
            \wheel
            \saddle
            \angletheta
            \draw[force] (0,1) -- ([shift={($- cos(20)*(110:1.5)$)}] 0,1);
            \draw[force] (0,1) -- node [above right = 0mm and -3mm] {$F_{W,g}$} ($(0,1) + cos(20)*sin(20)*(1.5,0)$);
            \draw[force] (0,1) -- ($(0,1) - cos(20)*cos(20)*(0,1.5)$);
        \end{tikzpicture}
    }
    \caption{\label{fig:gravity-pull-on-s} The Effect of Gravity on $S$}
\end{figure}

Now, we will factor in gravity. We do not need to consider the effect of gravity on $W$ -- it only affects the friction between $W$ and the ground, and we already assumed that $W$ does not slip. However, gravity also has an affect on $S$: If $S$ is not upright, it creates a torque on it. The situation is sketched in \cref{fig:gravity-pull-on-s}.

The gravitational pull of strength $F_g = m_s \cdot g$ acts on the center of mass of $S$. It can be split into two parts. One of them causes a torque $\tau_{S, g}$ with a moment arm of length $r_\mi{com}$. The other part pulls along the rod, towards the wheel. At the wheel, this force is split once again, into a part that is perpendicular to the ground and can be ignored, and a force $F_{W, g}$ that pushes $W$ to the right. By the geometry of the unicycle, we can compute the strength of these forces:
\begin{align}
    \tau_{S, g} &= m_s \cdot g \cdot \sin \theta \cdot r_\mi{com} \label{eqn:defining-tau-s-g} \\
    F_{W, g} &= m_s \cdot g \cdot \cos \theta \cdot \sin \theta \label{eqn:defining-f-w-g}
\end{align}

We now have derived two forces pushing on $W$ to the right: $F_{W, \mi{mot}}$ caused by the driver or motor and $F_{W, g}$ caused by gravity. We call the sum of both forces $F_W$:
\begin{equation}\label{eqn:defining-f-w}
    F_W = F_{W, \mi{mot}} + F_{W, g}
\end{equation}

While $F_W$ is the only force that acts on $W$ horizontally, this still is not enough to describe the motion of $W$, since $W$ and $S$ are connected and $S$ is free to rotate. We will analyze the effects of such a force in the next section.

\subsection{Splitting an External Linear Force on \texorpdfstring{$W$}{W} and \texorpdfstring{$S$}{S}}\label{sct:splitting-fw}

\begin{figure}\centering
    \subcaptionbox{\label{subfig:splitting-fw-1} $F_W$ acts on the Unicycle}[.33\linewidth]{
        \centering\begin{tikzpicture}
            \surface
            \wheel
            \saddle[60]
            \angletheta[60]
            \draw[force] (0,1) -- node [below left] {$F_W$} (1.5,1);
        \end{tikzpicture}
    }\subcaptionbox{\label{subfig:splitting-fw-2} $F_{W,2}$ and an Isolated $W$}[.33\linewidth]{
        \centering\begin{tikzpicture}
            \surface
            \wheel
            \draw[force] (0,1) -- node [below] {$F_{W,2}$} (0.5,1);
        \end{tikzpicture}
    }\subcaptionbox{\label{subfig:splitting-fw-3} $F_{W,1}$ and an Isolated $S$}[.33\linewidth]{
        \centering\begin{tikzpicture}
            \surface
            \draw[line width = 2mm, black!10] (-2,1) -- (2,1);
            \saddle[60]
            \angletheta[60]
            \draw[force] (0,1) -- node [below] {$F_{W,1}$} (1.0,1);
        \end{tikzpicture}
    }
    \caption{\label{fig:splitting-fw} Splitting an External Force}
\end{figure}
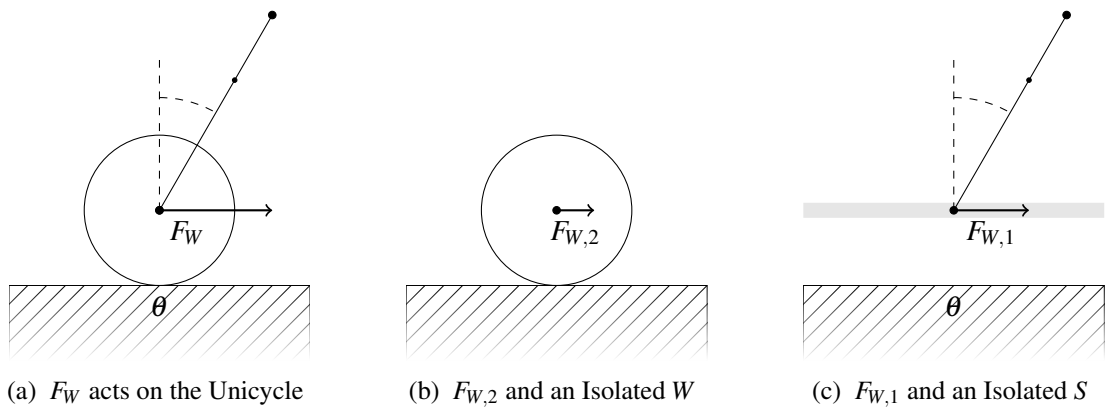

We will now derive the equations for a unicycle where an external force $F_W$ acts on the wheel as depicted in \cref{subfig:splitting-fw-1}. For this consideration, we will assume that no other forces act on the system.

We consider the effects on $W$ and on $S$ separately. The part of the force that only affects the wheel will be called $F_{W,2}$, the part that only affects $S$ will be called $F_{W,1}$. We know that
\begin{equation}
    F_W = F_{W,1} = F_{W,2}
\end{equation}
must hold. We consider the effect of $F_{W,2}$ on a wheel (see \cref{subfig:splitting-fw-2}) and the effect of $F_{W,1}$ on $S$ (see \cref{subfig:splitting-fw-3}) in isolation, assuming that the lower end of $S$ may not move up or down.

For the scenario in \cref{subfig:splitting-fw-2}, we can easily compute the acceleration of $W$:
\begin{equation}
    a_W = \frac{F_{W,2}}{m_W}
\end{equation}

In the scenario in \cref{subfig:splitting-fw-3}, the force $F_{W,1}$ is acting on an object that can both rotate around its center of mass and translate linearly. In this case, both things happen, and we can compute both the linear and angular acceleration:
\begin{align}
    a_S &= \frac{F_{W,1}}{m_S} \\
    \alpha_S &= \frac{r_\mi{com} \cdot F_{W,1} \cdot \cos \theta}{I_S}
\end{align}

Since the lower end of $S$ is fixed to $W$, we know that their acceleration is the same:
\begin{equation}
    a_S + \alpha_S \cdot r_\mi{com} \cdot \cos \theta = a_W
\end{equation}

From this, we immediately obtain the ratio of the two forces, which we will call $\beta$:
\begin{equation}\label{eqn:defining-beta}
    \beta \coloneqq \frac{F_{W,1}}{F_{W,2}} = \frac{m_S}{m_W} \cdot \round{1 + 3 \cos^2 \theta}
\end{equation}

This allows us to finally quantify both parts of $F_W$:
\begin{align}
    F_{W,1} &= F_W \cdot \frac{\beta}{1 + \beta} \label{defining-f-w-1} \\
    F_{W,2} &= F_W \cdot \frac{1}{1 + \beta} \label{defining-f-w-2}
\end{align}

\subsection{Torque on \texorpdfstring{$S$}{S} by Linear Acceleration}

The force $F_{W,1}$ accelerates the bottom end of $S$ to the right. This causes both linear acceleration which we can ignore since $S$ is fixed to $W$, and rotation around the center of mass. By the geometry of the unicycle, we can quantify this torque $\tau_{S,W}$:
\begin{equation}\label{defining-tau-s-w}
    \tau_{S,W} = F_{W,1} \cdot r_\mi{com} \cdot \cos \theta
\end{equation}

Therefore, we know all torques affecting $S$, and can compute the total torque on $S$:
\begin{equation}\label{eqn:defining-tau-s}
    \tau_S = -\tau_{W, \mi{mot}} + \tau_{S, g} + \tau_{S, W}
\end{equation}

\subsection{Flow Equations}

Now, we can finalize our model by specifying how the position $x_W$ of the wheel and the angle $\theta$ of the saddle evolve:
\begin{align}
    \dot{x_W} &= v_W \label{eqn:flow-xw} \\
    \dot{v_W} &= \frac{F_{W, 2}}{m_W} \label{eqn:flow-vw} \\
    \dot{\theta} &= \omega \label{eqn:flow-theta} \\
    \dot{\omega} &= \frac{\tau_S}{I_S} \label{eqn:flow-omega}
\end{align}

\end{document}